# Correlated metallic two-particle bound states in Wannier–Stark flatbands

Arindam Mallick,[1, *] Alexei Andreanov,[1, 2, †] and Sergej Flach[1, 2, ‡]

[1]*Center for Theoretical Physics of Complex Systems, Institute for Basic Science (IBS), Daejeon 34126, Republic of Korea*
[2]*Basic Science Program, Korea University of Science and Technology (UST), Daejeon 34113, Korea*
(Dated: April 27, 2022)

Tight-binding single-particle models on simple Bravais lattices in space dimension $d \geq 2$, when exposed to commensurate DC fields, result in the complete absence of transport due to the formation of Wannier–Stark flatbands [Phys. Rev. Res. 3, 013174 (2021)]. The single-particle states localize in a factorial manner, i.e., faster than exponential. Here, we introduce interaction among two such particles that partially lifts the localization and results in metallic two-particle bound states that propagate in the directions perpendicular to the DC field. We demonstrate this effect using a square lattice with Hubbard interaction. We apply perturbation theory in the regime of interaction strength $(U) \ll$ hopping strength $(t) \ll$ field strength $(\mathcal{F})$, and obtain estimates for the group velocity of the bound states in the direction perpendicular to the field. The two-particle group velocity scales as $U(t/\mathcal{F})^\nu$. We calculate the dependence of the exponent $\nu$ on the DC field direction and on the dominant two-particle configurations related to the choices of unperturbed flatbands. Numerical simulations confirm our predictions from the perturbative analysis.

## I. INTRODUCTION

Flatbands [1–4] are macroscopically degenerate dispersionless bands in the band structure of single particle lattice systems. The degeneracy makes them highly susceptible to perturbations, and they show various interesting phenomena in the presence of disorder [5, 6], nonlinearity [7, 8] and interactions [9–11]. In particular, flatband systems with interaction were suggested as candidates for high temperature superconductors [12, 13]. Translationally invariant flatband systems with short-range hopping are known to support compact localized eigenstates (CLSs) [14]. Flatbands have also been observed experimentally in various condensed matter systems [15, 16] and have been suggested as a useful platform to develop quantum technologies [17–19].

The energy spectrum from a single-particle tight-binding Hamiltonian on an infinite Bravais lattice is given by one single dispersive band. When exposed to a commensurate DC field (i.e., the direction perpendicular to the field is a lattice vector) of strength $\mathcal{F}$ the spectrum turns into an equidistant ladder of Wannier–Stark (WS) flatbands [20, 21], provided there is no equipotential hopping. The absence of dispersion in any of the bands implies zero group velocity and therefore complete particle localization. These flatbands do not support CLS, essentially because there is an infinite number of flatbands. Their eigenvectors are factorially (or super-exponentially) localized in real lattice space [21].

For fine-tuned flatband Hamiltonians featuring CLSs, interactions among fermions or bosons typically induce motion in one-dimensional lattices [22, 23]. Finetuned interactions can protect particle localization either partially or completely [24], making such interacting systems candidates for many-body localization [25, 26]. Delocalization can also be induced in translationally invariant flatband models by interactions in the presence of magnetic fields [27–30].

Here we address the following question: how will flatbands without CLSs react to perturbations? In particular, how will interactions alter the localization properties and lead to transport? To address this issue, we study the the effect of contact interaction between two particles in the WS flatband setup. With no interaction, particles follow the zero dispersion of the bands and can occupy eigenstates from the same or different single-particle WS flatbands. Hubbard-like onsite interaction will locally disrupt the flatness and induce nonzero transport. The contribution of two interacting particles to the transport depends on the total energy of unperturbed flatbands occupied by two non-interacting particles. We identified two quantitatively different cases. The particles form a bounded pair and move only along the direction perpendicular to the field. The formation of the bounded pair follows from the observation that widely separated particles do not feel the interaction and are therefore localized, as in the single-particle case. Motion along the field direction is prohibited due to energy conservation. As a consequence of the interaction, the particles acquire a finite group velocity and propagate ballistically in the direction perpendicular to the field.

We use degenerate perturbation theory with interaction strength $U$ to analyze the problem. We further simplify the problem by approximating the eigenstates to a CLS in the limit of weak hopping. This allows us to estimate the group velocity associated with the center of mass of two distinguishable particles as a function of hopping, interaction, and DC field strength.

In Section II we set the stage by introducing the single-particle Hamiltonian and its properties. Section III discusses the interacting Hamiltonian for two particles, either distinguishable or spinless bosons, or fermions with opposite spin. We use a geometrical approach to analyze the two-particle motion in the presence of the interaction in Section IV. Section V gives details of the perturbative calculation of the group velocity, and Section VI provides the corresponding numerical results. Discussion and conclusions follow in Section VII.

* marindam@ibs.re.kr
† aalexei@ibs.re.kr
‡ sflach@ibs.re.kr

## II. SETTING THE BACKGROUND

We label lattice sites of an infinite square lattice by a pair of integers $(n, m)$ and consider a single-particle tight-binding Hamiltonian with homogeneous hopping $t$ and uniform DC field $\vec{\mathcal{E}} = \mathcal{F}(x, y)$,

$$\hat{\mathcal{H}} = \sum_{n,m} \left[ \vec{\mathcal{E}} \cdot (n, m) \hat{b}_{n,m}^\dagger \hat{b}_{n,m} - t \sum_{i,j} \hat{b}_{n-i,m-j}^\dagger \hat{b}_{n,m} \right]. \quad (1)$$

$$\mathcal{F} = |\vec{\mathcal{E}}| / \sqrt{x^2 + y^2}, \qquad xy \neq 0.$$

Here, $\hat{b}_{n,m}^\dagger$ and $\hat{b}_{n,m}$ are creation and annihilation operators at site $(n, m)$, respectively. The indices $(i, j)$ take values $\{(0, \pm 1), (\pm 1, 0)\}$ for the nearest-neighbor hopping considered in this work. The components of the field direction $x$ and $y$ are taken as mutually prime integers. This ensures that the field has a commensurate field direction [21], i.e., translational invariance along the direction perpendicular to the field. We introduce coordinates along and perpendicular to the field direction [21],

$$z = nx + my, \qquad w = ny - mx. \quad (2)$$

Note that $z$ is a scaled version of the field coordinate $\vec{\mathcal{E}} \cdot (n, m) / |\vec{\mathcal{E}}| = (nx + my) / \sqrt{x^2 + y^2}$ in units of lattice spacing where $n$ or $m$ changes by 1, and similarly $w$ is a scaled version of $(ny - mx) / \sqrt{x^2 + y^2}$. For a square lattice, $z$ takes all possible integer values while $w$ can be decomposed into

$$w = w_0(z) + (x^2 + y^2)\eta, \quad (3)$$

with a $z$-dependent integer part $w_0(z) = z(\tau_2 y - \tau_1 x)$ and an integer $\eta$. Here, $\tau_1$ and $\tau_2$ are two integers that satisfy $\tau_1 y + \tau_2 x = 1$. Since translational invariance is preserved along the direction orthogonal to the field, i.e. for $\eta$, the Hamiltonian $\hat{\mathcal{H}}$ is block diagonal in one-dimensional Bloch momentum space conjugated with $\eta$. For our choice of Hamiltonian parameters, this leads to a set of equidistant WS flatbands $E_a = a\mathcal{F}$ with band index $a \in \mathbb{Z}$ [21]. Each flatband $E_a$ has a complete orthogonal set of eigenstates ($J_\mu$ is the order $\mu$ Bessel function of the first kind) given by

$$|\Phi(a, l)\rangle = \sum_{n,m \in \mathbb{Z}} J_{n-n_0}\left(\frac{2t}{x\mathcal{F}}\right) J_{m-m_0}\left(\frac{2t}{y\mathcal{F}}\right) |z, \eta\rangle$$

$$a = n_0 x + m_0 y, \quad l = \tau_1 n_0 - \tau_2 m_0, \quad (4)$$

that are super-exponentially localized in real space [21]. The value of field coordinate $z$ coincides with the band index $a$, and $l$ equals the value of $\eta$ at the site $(n_0, m_0)$. The integer $l$ also labels the eigenstates and corresponds to the maximum weight of $|\Phi(a, l)\rangle$ along $\eta$ [see Appendix A for details]. The ratio $|t/\mathcal{F}|$ controls the spreading of the eigenfunction. In the limit $|t/\mathcal{F}| \to 0$, the particle is localized at site $(a, l) \equiv (n = n_0, m = m_0)$ of the square lattice, giving a CLS that occupies a single site. For $|t/\mathcal{F}| \ll 1$, one can construct a CLS that approximates $|\Phi(a, l)\rangle$ based on a truncation of the Taylor expansion of the Bessel function:

$$J_\mu(2s) = s^{|\mu|} [\text{sign}(\mu)]^\mu \left( \frac{1}{|\mu|!} - \frac{1}{(1 + |\mu|)!} s^2 \right.$$
$$\left. + \frac{1}{2(2 + |\mu|)!} s^4 + \dots \right). \quad (5)$$

We use these approximate CLSs in the analysis that follows, since they are easier to handle than the full super-exponentially localized eigenfunctions.

## III. THE MODEL

We consider two distinguishable particles with annihilation (creation) operators $\hat{b}_{1,nm}$ ($\hat{b}_{1,nm}^\dagger$) and $\hat{b}_{2,nm}$ ($\hat{b}_{2,nm}^\dagger$) at site $(n, m)$. For indistinguishable bosons, $\hat{b}_1 = \hat{b}_2$, and they belong to the same Hilbert space and follow the usual bosonic commutation relations. In what follows, we focus on the case of distinguishable particles for convenience. In the presence of Hubbard-like contact interaction, we define the following two-particle Hamiltonian on a square lattice as

$$\hat{\mathcal{H}}_{12} = \hat{\mathcal{H}}_1 + \hat{\mathcal{H}}_2 + \hat{V}, \quad (6)$$

$$\hat{V} = U \sum_{n,m} \hat{b}_{1,nm}^\dagger \hat{b}_{1,nm} \hat{b}_{2,nm}^\dagger \hat{b}_{2,nm}, \quad (7)$$

where $U$ is the interaction strength. $\hat{\mathcal{H}}_{1,2}$ are the single-particle Hamiltonians as in Eq. (1) with $\hat{b}$ replaced by $\hat{b}_{1,2}$. The hopping strength $t$ is the same for both particles.

The spectrum of $\hat{\mathcal{H}}$ in Eq. (1) is composed of flatbands $E(k) = a\mathcal{F}, a \in \mathbb{Z}$, and therefore, the spectrum of $\hat{\mathcal{H}}_{12}$ for $U = 0$ is given by a tower of flatbands $E(k_1, k_2) = E = \mathcal{F}(a_1 + a_2)$, where the particles fill single-particle flatbands with indices $a_1, a_2$. Depending on the values of $a_1$ and $a_2$, the total eigenenergy $E$ is either odd or even (in units of $\mathcal{F}$). We refer to these two cases as "O-band" and "E-band", respectively. These two types of particle configurations give different contributions to the transport in the presence of interaction as we show below.

We define the center of mass (c.m.) and relative coordinates for $z, w$ [Eq. (2)] of the two particles as

$$z_c = \frac{z_1 + z_2}{2}, \qquad z_r = \frac{z_1 - z_2}{2},$$

$$w_c = \frac{w_1 + w_2}{2}, \qquad w_r = \frac{w_1 - w_2}{2}, \quad (8)$$

so that we can rewrite the interaction matrix as follows:

$$\hat{V} = U \sum_{z_c, w_c, z_r, w_r} \delta_{z_r, 0} \delta_{w_r, 0} |z_c, w_c, z_r, w_r \rangle \langle z_c, w_c, z_r, w_r|. \tag{9}$$

This acts as an identity in the $(z_c, w_c)$ space and as a single-site defect in the $(z_r, w_r)$ space. Such interaction maintains the translational invariance along the $w_c$ coordinate present in the non-interacting case. Therefore, $\mathcal{H}_{12}$ is block diagonal in the c.m. momentum $k$-space conjugated to $w_c$, while the matrix elements in each block depend on three coordinates, $(z_c, z_r, w_r)$. For $U \neq 0$, the eigenvalues in this new coordinate system are functions of the Bloch momentum $k$ and can be dispersive, implying transport.

## IV. GEOMETRIC ANALYSIS

The single-particle eigenstates $|\Phi(a, l)\rangle$ (4) are labeled by the pair of integers $a, l$ that are related to the field coordinates $z, \eta$ and can be organized into a square lattice. In this representation, the site $(a, l)$ of this square lattice belongs to the single-particle flatband of index $a$ for all $l$. For convenience, we refer to these lattice sites as "$F$-sites", while the lattice sites in the real space square lattice are referred to simply as "sites". The eigenstates $|\Phi(a, l)\rangle$ are not compact but super-exponentially localized, and therefore a particle localized at a single F-site has nonzero amplitudes on all sites. However, every F-site can be approximated by a CLS in real space, using the truncated expansion Eq. (5) in $t/\mathcal{F}$ in the exact eigenstate from Eq. (4). This is an important observation for the derivation of the transport properties that is provided below.

For two particles, we define the relative and c.m. coordinates along and perpendicular to the DC field for the F-sites $(a_1, l_1)$ and $(a_2, l_2)$ by

$$\xi = l_1 - l_2, l_1 + l_2 \equiv 2l_c + \xi, \text{ where } l_c, \xi \in \mathbb{Z}, \tag{10}$$

similarly to Eq. (8). The two particle state

$$|\Phi(a_1, a_2, l_c, \xi)\rangle := |\Phi(a_1, l_1)\rangle \otimes |\Phi(a_2, l_2)\rangle \tag{11}$$

is now parameterized by the four coordinates $a_1, a_2, \xi, l_c$. As discussed above, we expect to see transport of a bound pair along the center of mass of the two particles, i.e., the $l_c$ coordinate. In perturbation theory, the leading contribution to interaction comes from the spatial overlap between two-particle Fock states of the non-interacting problem with the same eigenenergy given by

$$E^{(0)} = (a_1 + a_2)\mathcal{F}. \tag{12}$$

In the F-site representation, the hopping terms in Eq. (6) are diagonal and act as an onsite potential, while interaction contains both diagonal and off-diagonal terms and

induces hopping between the F-sites:

$$\hat{\mathcal{H}}_{12} = \sum \mathcal{F}(a_1 + a_2) |\Phi(\mathbf{n})\rangle\langle\Phi(\mathbf{n})| + \hat{V},$$

$$\hat{V} = U \sum_{n, m} \hat{b}^\dagger_{1, nm} \hat{b}_{1, nm} \hat{b}^\dagger_{2, nm} \hat{b}_{2, nm}$$

$$= \sum_{\mathbf{n}, \mathbf{n}'} |\Phi(\mathbf{n}')\rangle \langle\Phi(\mathbf{n}')|\hat{V}|\Phi(\mathbf{n})\rangle \langle\Phi(\mathbf{n})|, \tag{13}$$

where $\mathbf{n} = (a_1, a_2, l_c, \xi)$ and $\mathbf{n}' = (a'_1, a'_2, l'_c, \xi')$. Since F-states are super-exponentially localized in real space rather than compact, the interaction induces hopping between all F-sites, even if most of the hoppings between F-sites are super-exponentially suppressed. In the presence of the interaction, the two-particle states [Eq. (11)] are no longer eigenstates of the Hamiltonian of Eq. (6). In the limit $|t/\mathcal{F}| \to 0$, the eigenfunction $|\Phi(a, l)\rangle$, i.e., the F-site, occupies a single site. In this case, the two-particle eigenstates remain single-site CLSs even in the presence of interaction, and the non-interacting eigenenergies are merely shifted by the interaction $E_U = E^{(0)} + U$, and thus the macroscopic degeneracy survives and particle transport is still prohibited. Away from $t/\mathcal{F} = 0$, other corrections are generated to the two-particle eigenenergies $E_U = E^{(0)} + E^{(1)} + \ldots$ that can induce transport. We seek to understand how the transport emerges and how the velocity of the two-particle motion scales with $t/\mathcal{F}$ with first-order perturbation in interaction strength $U$. Since the $U = 0$ case corresponds to flatbands, implying macroscopically degenerate eigenstates, we use the degenerate perturbation theory for the eigenenergies. The first order correction to the flatband eigenenergy $E^{(1)}$ is given by the eigenvalues of $\hat{V}^{(1)}$:

$$\hat{V}^{(1)} = \sum_{\mathbf{n}, \mathbf{n}'} \delta_{a_1 + a_2, a'_1 + a'_2} |\Phi(\mathbf{n}')\rangle \langle\Phi(\mathbf{n}')|\hat{V}|\Phi(\mathbf{n})\rangle \langle\Phi(\mathbf{n})|. \tag{14}$$

Here, $\hat{V}^{(1)}$ is the part of the interaction matrix $\hat{V}$ subject to the energy conservation constraints enforced by the Kronecker symbol $\delta_{a_1 + a_2, a'_1 + a'_2}$. Every matrix element $\langle\Phi(\mathbf{n}')|\hat{V}|\Phi(\mathbf{n})\rangle$ in real lattice space is given by the overlap between non-interacting two-particle eigenstates $|\Phi(a_1, a_2, l_c, \xi)\rangle$ and $|\Phi(a'_1, a'_2, l'_c, \xi')\rangle$. These eigenstates are not compact but super-exponentially localized in real space, and therefore the matrix elements are expected to be nonzero even if super-exponentially small for all sites. It is important to note the translational invariance with respect to $l_c, l'_c$ that is preserved even in the presence of interaction, which is the same translational invariance that we discussed above following Eq. (9). This translational invariance allows us to block diagonalize $\hat{V}^{(1)}$ and define a respective momentum $k$, conjugated to $l_c, l'_c$. The remaining problem is to diagonalize $\hat{V}^{(1)}$ for a given $k$, and check the dispersion of the eigenvalues $E^{(1)}(k)$. Yet the analytical calculation of $E^{(1)}(k)$ is still challenging. Energy conservation splits the partially diagonalized blocks into sub-blocks of E- and O-bands. The corresponding sub-blocks are still of infinite size in parameters $a_1, a_2, a_1', a_2', \xi, \xi'$.

As a consequence, we have to resort to further approximations and use compact approximations of the F-states in powers of $t/\mathcal{F}$, that we discussed above with Eqs. (4,5), to compute $E^{(1)}$ as a function of $t/\mathcal{F}$ to the leading order. We need to identify the two-particle configurations that give the leading corrections, in parameter $t/\mathcal{F}$, to the $t/\mathcal{F} = 0$ result. In what follows, we refer to this simply as the leading term or correction. For this, a geometric analysis of the non-interacting ($U = 0$) two-particle eigenstates is necessary. In the following subsections, we consider the case of the field direction $(x, y) = (1, 1)$ and demonstrate that the contributions of the non-interacting two-particle eigenstates $|\Phi\rangle$—namely, those from the E- or O-band—are different, and that the states $|\Phi\rangle$ have to be expanded at least up to order $(t/\mathcal{F})^2$ for this particular field direction.

Before we proceed, it is useful to introduce some conventions. Since the F-states are parameterized by the flatband index $a$ and the location (in the flatband) $l$, the F-states corresponding to a given flatband $a$ form lines of F-sites, as illustrated by the black lines containing red spheres in Fig. 1. This mapping of the F-sites onto lattice sites helps to analyze the overlapping of the eigenfunctions (in real space), which is crucial for finding the leading correction and understanding the effect of interaction as well as the emergence of particle transport.

### A. $U = 0$ E-bands

The $U = 0$ E-bands are formed by pairs of particles occupying F-sites $(n_1, m_1, n_2, m_2) = (n + \xi_e, m + \xi_e', n + \sigma_e, m + \sigma_e')$ with integers $\xi_e, \xi_e', \sigma_e, \sigma_e'$ and have eigenenergy $E_a^{(0)} = 2a\mathcal{F}$. The values of $\xi_e, \xi_e', \sigma_e, \sigma_e'$ are constrained by the energy conservation, i.e., the $z_1 + z_2$ value equal to $n_1 x + m_1 y + n_2 x + m_2 y = 2(nx + my) = 2a$ should not change, imposing the constraint $\xi_e x + \xi_e' y = -(\sigma_e x + \sigma_e' y)$. Three following possible cases are shown in Fig. 1 for the field direction $x = y = 1$.

1. Both particles occupy the same F-site (the same single-particle flatband). The values $\xi_e = \sigma_e = \beta_e y, \xi_e' = \sigma_e' = -\beta_e x$ are parameterized by a single integer, $\beta_e$, that encodes the position of the F-site in the band $E_a^{(0)}$. These F-states are shown in Fig. 1 within dash-dotted black circles.

2. The particles are located symmetrically with respect to the flatband $a$ in bands $a \pm 1$, occupying different F-sites along the field. This gives the constraint $\xi_e y - \xi_e' x = \sigma_e y - \sigma_e' x$. This case is shown in Fig. 1 with yellow-shaded ellipses, for which $(\xi_e, \xi_e', \sigma_e, \sigma_e') = \pm(1, 0, 0, -1)$.

3. The two particles form a horizontal or vertical pair of F-states, as indicated by violet dashed ellipses

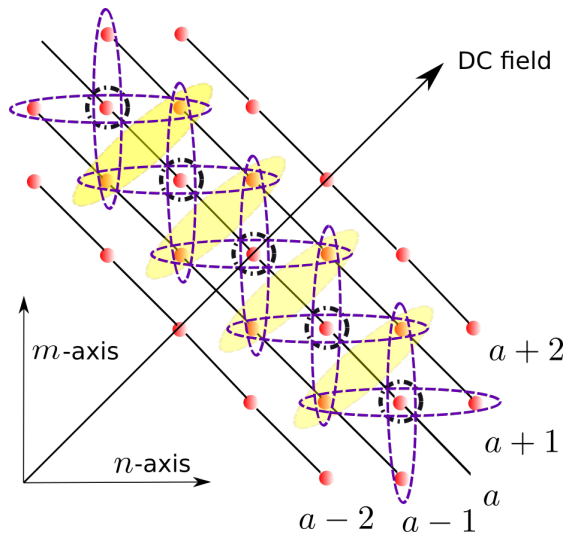

FIG. 1. Schematic representation of the eigenstate pairs for E-bands with eigenenergies $2a\mathcal{F}$ on the square lattice formed by F-sites for the DC field direction $(x, y) = (1, 1)$. Each of the lattice sites (red spheres) is a single particle F-state $|\Phi(a, l)\rangle$. $a \in \mathbb{Z}$ simultaneously denotes the single-particle band index and the $z$ coordinate. The constant $a$ lines are drawn over F-states $|\Phi(a, l)\rangle$ for different $l$, which characterize the average values of the perpendicular coordinate $\eta$ for fixed $a$.

connecting flatbands $a \pm 1$ in Fig. 1. The constraints on the coordinates are $\xi_e = -\sigma_e = \zeta_e, \xi_e' = -\sigma_e' = \zeta_e'$, and $\zeta_e, \zeta_e' \in \mathbb{Z}$ with $\zeta_e$ and $\zeta_e'$ not being zero simultaneously. In this setting, one particle is in the single-particle band $a + x\zeta_e + y\zeta_e'$ and the other particle is in the single-particle band $a - x\zeta_e - y\zeta_e'$.

If we approximate the two-particle state $|\Phi(a_1, a_2, l_c, \xi)\rangle$ by a CLS in powers of $t/\mathcal{F}$ using Eq. (5), then these three cases produce different leading corrections in powers of $t/\mathcal{F}$. These corrections correspond to the overlaps between the F-sites. For the first case, the contribution is $(t/\mathcal{F})^0$, while the second and third cases both contribute as $(t/\mathcal{F})^2$. These pairs of the F-sites form a quasi-one-dimensional chain along the direction perpendicular to the field, if shifted along the flatband $a$, i.e., along $l_c, \xi$ for the two-particle states. Such connected chains only appear at leading order $(t/\mathcal{F})^2$ of the expansion of matrix elements of $\hat{V}^{(1)}$ [using Eq. (5)], and provide a hopping mechanism for pairs of particles and therefore transport. All the other pairs not shown in the figure give either $(t/\mathcal{F})^2$ or higher powers of $t/\mathcal{F}$ as sub-leading corrections.

### B. $U = 0$ O-bands

For the O-bands, the two particles cannot occupy F-sites in the same flatband. Therefore, they have to occupy F-sites in neighboring flatbands, or flatbands separated by an even number of flatbands to ensure an odd total energy $E_a^{(0)} = (2a-1)\mathcal{F}$. The O-bands are formed by different types of such pairs of particles. The coordinates of the F-sites are given by $(n_1, m_1, n_2, m_2) = (n, m, n - \alpha_0\tau_2 + \beta_o y, m - \alpha_0\tau_1 - \beta_o x)$ for odd integer $\alpha_o$ and integer $\beta_o$ that control the position of the pair along the flatband. The $\tau_1, \tau_2$ are integers satisfying $\tau_1 y + \tau_2 x = 1$. There are many solutions to this relation, but the specific choice does not affect any physical observables and is rather made out of convenience [21]. For our analysis we choose $\tau_1 = 0, \tau_2 = 1$. There are several types of pairs; we list the most important ones here and depict them in Fig. 2.

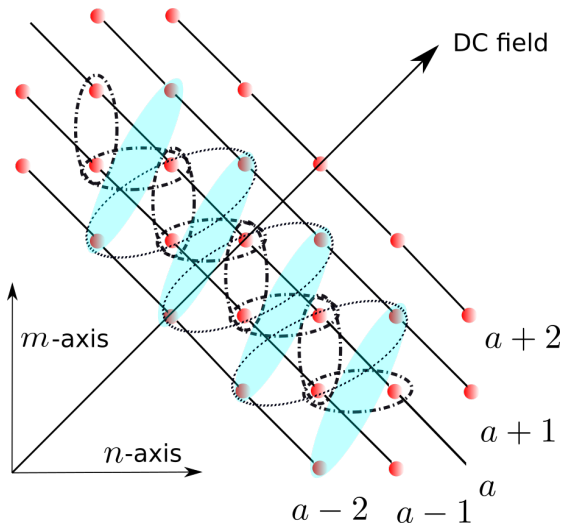

FIG. 2. Schematic representation of the eigenstate pairs for O-bands with eigenenergies $(2a-1)\mathcal{F}$ on the square lattice of F-sites. The setup is similar to Fig. 1.

1. The two particles can form a pair of nearest neighbor F-sites at bands $a$ and $a-1$, either vertically along the $m$-axis obtained by setting $\alpha_o = 1, \beta_o = 1$ or horizontally along the $n$-axis obtained by setting $\alpha_o = 1, \beta_o = 0$. These are depicted as dash-dotted black ellipses in Fig. 2.

2. The other pair can be formed by the particles located at the F-sites roughly oriented along the diagonal direction, as shown in Fig. 2 by shaded cyan ellipses obtained by setting $\alpha_o = 3, \beta_o = 2$, and by dotted black ellipses obtained by setting $\alpha_o = 3, \beta_o = 1$. They are located at bands $a+1$ and $a-2$.

Expanding the two-particle F-states $|\Phi(a_1, a_2, l_c, \xi)\rangle$ in powers of $t/\mathcal{F}$ using Eq. (5), we find the leading terms of order $(t/\mathcal{F})^1$ or higher. Therefore, the matrix elements appearing in $\hat{V}^{(1)}$ (14) are at least of order $(t/\mathcal{F})^2$ or higher. Similar to the E-band cases, the pairs also form different types of quasi one dimensional chains along the direction perpendicular to the field.

Following a similar approach, we discuss the case for another field direction in Appendix B, when either $x$ or $y$

is even (the case in which both are even is excluded since we assumed $x, y$ to be mutually prime).

## C. Contributions from E- and O-bands in pair transport for $U \neq 0$

Since the F-state pairs depicted in Figs. 1 and 2 overlap in the real lattice space, they produce nonzero off-diagonal matrix elements of $\hat{V}^{(1)}$ and give rise to a nonzero contribution to the energy shift from $E^{(0)}$. This shift is in turn responsible for the nonzero group velocity of the particles and hence the transport. As discussed above, it is challenging to diagonalize the matrix $\hat{V}^{(1)}$ analytically, even after the partial block diagonalization, and so we resort to expansion in powers of $t/\mathcal{F}$ generated by Eq. (5). The diagonal elements of the matrix $\hat{V}^{(1)}$ correspond to the overlap between the same pairs of particles, and the leading order term is $(t/\mathcal{F})^\mu$, where $\mu$ is a positive integer that depends on the distance between the F-sites within the pair. The off-diagonal terms involve two different pairs from the quasi one dimensional chains discussed above. These chains can be formed by pairs of the same or different types, and their contributions would produce terms with different leading powers of $t/\mathcal{F}$. We computed the leading order contribution $(t/\mathcal{F})^\nu$ for certain field directions: for the $(1,1)$ direction, we found $\nu = 2$, and for the $(2,-3)$ field direction, we found $\nu = 5$ [see Appendix B]. In these examples, we found that the leading contribution comes from the O-bands rather than the E-bands. Based on these results, we make the following conjecture:

*Conjecture*— The leading order term for the off-diagonal elements of the interaction matrix $\hat{V}^{(1)}$ scales as $|t/\mathcal{F}|^{|x|+|y|}$.

Therefore, if we consider the energy correction up to the first order in $U$, we expect a bounded pair motion through the quasi one dimensional chains depicted in Figs. 1 and 2, along the direction perpendicular to the field. The corresponding group velocity

$$v_g \propto (\partial E^{(1)}/\partial k) \propto |U||t/\mathcal{F}|^\nu, \qquad (15)$$

where $\nu \geq |x| + |y|$.

## V. PERTURBATIVE CALCULATION

We are now ready to use the first-order degenerate perturbation in $U$ as given by Eq. (14), and diagonalize $\hat{V}^{(1)}$ using the leading order contributions in $t/\mathcal{F}$ to its matrix elements in order to compute the eigenvalues of $\hat{V}^{(1)}$ and analyze the particle transport. Our perturbative analysis is valid for $U \ll t \ll \mathcal{F}$; the approximation of the super-exponentially localized eigenstates $|\Phi(a, l)\rangle$ by CLSs is justified for $t \ll \mathcal{F}$. We derive the analytical results for the field direction $(x, y) = (1, 1)$ and we choose $\tau_1 = 0, \tau_2 = 1$. Using the series expansion of the F-states generated by Eq. (5) up to a finite order in $t/\mathcal{F}$, we compute the eigenvalues $E^{(1)}$ of $\hat{V}^{(1)}$. Some of them acquire finite dispersion with respect to the center of mass momentum $k$ (conjugated to the center of mass coordinate $l_c$ defined in Eq. (10)) if we take into account the contributions to the interaction matrix elements at least up to order $(t/\mathcal{F})^2$ (as follows from the geometrical arguments discussed above). In this case, the interaction matrix $\hat{V}^{(1)}$ has two dispersive eigenvalues,

$$
\begin{aligned}
E_{\mathrm{e}}^{(1)} &= \frac{U}{2}\left[1 - 8(t/\mathcal{F})^2\right] \\
&\quad \mp \frac{U}{2}\sqrt{1 - 16(t/\mathcal{F})^2 + (t/\mathcal{F})^4(32\cos k + 80)}, \\
E_{\mathrm{o}}^{(1)} &= 4U(t/\mathcal{F})^2\left(1 \pm \cos\frac{k}{2}\right),
\end{aligned}
\tag{16}
$$

for E- and O-bands, respectively. Therefore, we find two-particle dispersive bands emerging due to interaction, and the particles acquire a nonzero group velocity in the c.m. coordinate $l_c$. The group velocities corresponding to the different energy bands (E- and O-bands respectively) are

$$
\begin{aligned}
v_{g,\mathrm{e}} &= \frac{\partial E_{\mathrm{e}}^{(1)}}{\partial k} = \mp 8U(t/\mathcal{F})^4\sin k + \mathcal{O}((t/\mathcal{F})^6), \\
v_{g,\mathrm{o}} &= \frac{\partial E_{\mathrm{o}}^{(1)}}{\partial k} = \mp 2U(t/\mathcal{F})^2\sin\frac{k}{2} .
\end{aligned}
\tag{17}
$$

The group velocity for the O-bands $\mathcal{O}((t/\mathcal{F})^2)$ is larger compared to that of the E-bands $\mathcal{O}((t/\mathcal{F})^4)$. As a result, within the first order perturbation in $U$ and second order approximation in $t/\mathcal{F}$, the particles form a bounded pair that can move in the direction perpendicular to the field. It also instructive to point out that higher derivatives of $E^{(1)}(k)$, including the second derivative responsible for the wave packet spreading, display the same scaling with $t/\mathcal{F}$ as the group velocity, as follows directly from Eq. (16). All the other eigenvalues remain flat at this order $\mathcal{O}((t/\mathcal{F})^2)$, and therefore have zero group velocity and the higher order derivatives with respect to $k$.

## VI. NUMERICAL RESULTS

We numerically checked our perturbation results for the group velocity of the two-particle cases shown above. For this, we followed the unitary evolution of the wave function, starting from an initial two-particle wave packet,

$$
|\psi(T)\rangle = e^{-i\hat{\mathcal{H}}_{12}T}\,|\psi(T=0)\rangle .
\tag{18}
$$

We use initial states $|\psi(0)\rangle$ localized at the center of the square lattice. Therefore their Fourier series contains all momenta $k$ with roughly the same Fourier coefficients. As a result given Eq. (17) we expect to see negligible variations in the first moment of the coordinate $w_c$. However the variation of the second moment is expected to be significant and follow the same scaling as the group velocities.

The evolution is implemented using SciPy [31], avoiding the full diagonalization of the Hamiltonian and based on a combination of scaling and squaring methods, with a finite truncation of the Taylor's series of $e^{-i\hat{\mathcal{H}}_{12}T}$ (see the algorithm 5.2 in Ref. [32] for details). To ensure the correctness of our results, we checked that this propagation scheme produces relative errors less than $10^{-10}$ for both the total norm and the total energy. To capture the distinct features of the E- and O-bands, we used two different localized initial states $|\psi(0)\rangle$ that are eigenstates at $U = 0$:

- for E-bands $|\Phi(a_1, l_1)\rangle \otimes |\Phi(a_2, l_2)\rangle$,

- for O-bands $|\Phi(a_1, l_1)\rangle \otimes |\Phi(a_2 + 1, l_2)\rangle$,

where $a_1 = a_2 = 40$, $l_1 = l_2 = -20$. Both states are approximated up to first order in $t/\mathcal{F}$ so that they form a CLS on the real lattice sites. We put them at the center of a finite square lattice of size $41 \times 41$. The hopping strength is set to $t = 1$ and the time-evolution of these two-particle wave functions is followed up to time $T = 6000$ in units of hopping. In order to analyze the transport properties of the particles, we computed the time-evolution of the first and second moments of two-particle lattice site coordinates $z_c, z_r, w_c$ and $w_r$ as functions of $T$ and $\mathcal{F}$. We set the interaction strength to $U = 0.1 \ll 1.0$, and the field direction is taken along the diagonal of the square lattice, $(x, y) = (1, 1)$. To quantify the particle transport, we compute the average and the root mean square (RMS) values of the coordinates $w_c, z_c, z_r, w_r$ in lattice units, where the original coordinate $n$ or $m$ increments by 1,

$$
\begin{aligned}
\langle p \rangle, \mathrm{RMS} &= \sqrt{\langle p^2 \rangle - \langle p \rangle^2}, \\
(\sqrt{x^2 + y^2})p &= w_c, z_c, z_r, w_r,
\end{aligned}
\tag{19}
$$

as a function of time $T$. The factor $\sqrt{x^2 + y^2}$ is introduced to match the increments of the variables $\equiv (w_c, z_c, z_r, w_r)$ in units of lattice site coordinate [see the discussion around Eq. (2)] [21]. Based on the simple arguments and the above perturbative calculation, we anticipate no motion along $z_c, z_r$ where motion is suppressed by the field gradient, and a bounded value for $w_r$—the relative distance of the particles in the direction perpendicular to the field—where too much of a separation between the particles eliminates the effect of interaction and reinstates single particle localization. Figure 3 shows the results for the O-band initial state. The plots in the bottom inset in Fig. 3 confirm the linear (ballistic) increase with time $T$ of the center of mass coordinate $w_c$, while the other three coordinates $z_r, z_c, w_r$ saturate quickly and remain localized after a short initial transient. On the other hand, the first moment of $w_c$ shows a linear change with time. The weakness of the change is explained by the strong field value required to achieve the regime where our perturbative results are valid, $t \ll \mathcal{F}$. The top inset depicts the final time probability distribution for one of the two particles on the square lattice, i.e., after integrating out the other particle. Again as anticipated, the particle only has a significant presence along the direction perpendicular to the field, corresponding to motion perpendicular to the field. We find a qualitatively similar behavior for the case of the E-band initial state.

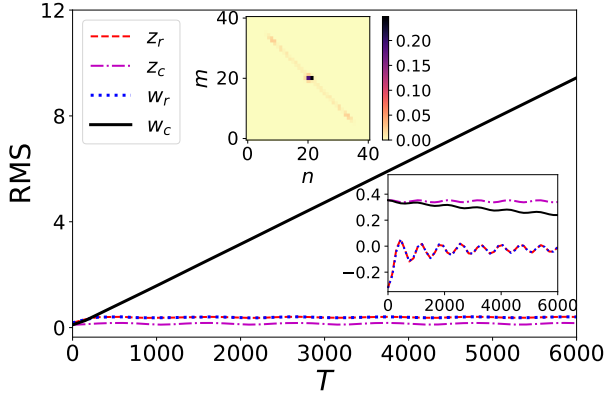

FIG. 3. Time evolution of the root mean square of position coordinates $w_c, z_c, w_r, z_r$ for the O-band initial state. The field direction is $(1,1)$ and the field strength is $\mathcal{F} = 9$. The physical length unit defined in Eq. (19) is used. Only the RMS of $w_c$ shows a linear increase while the RMS of all the other coordinates is bounded, indicating localization along these coordinates. The top inset shows the probability distribution of the second particle wave function at the final time of simulation. The probability is localized along the direction perpendicular to the field. A similar result is observed for the first particle. The bottom inset shows the first moments of the coordinates (as in the legend) as a function of time, with the same line styles and colors as in the main plot. Note that, the value of the first moment of $z_c$ is considerably larger than the first moments of the other coordinates. For convenience, $z_c - 20\sqrt{2}$ is plotted as the first moment of $z_c$ in the inset.

To confirm the predicted scaling of the group velocity as in Eq. (15), it is convenient to look at the RMS values that show a more pronounced variation compared to the first moment even for the strong fields considered. As we see in the main plot of Fig. 3, the RMS of the variables $z_c, z_r, w_r$ saturate with time, while only the RMS of $w_c$ displays a steady linear increase. We denote the RMS for $w_c$ as $Rw_c$ and the velocity of its increase as $Rv_c$. The latter is computed as a discrete derivative:

$$Rv_c(T, \mathcal{F}) = \frac{Rw_c(T + \Delta T, \mathcal{F}) - Rw_c(T - \Delta T, \mathcal{F})}{2\Delta T}, \quad (20)$$

for the field strengths $\mathcal{F} \in \{8, 9, \ldots, 16\}$ and 41 equally spaced discrete time-steps from $T = 0$ to $T = 6000$. Figure 4 shows $Rv_c$ as a function of time and field strength for the O-band initial state. The E-band intial state gives

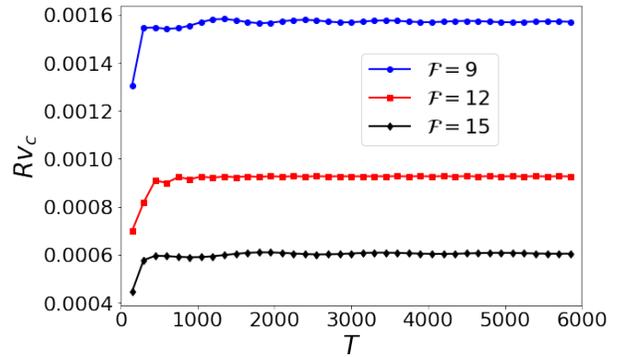

FIG. 4. RMS velocity $Rv_c$ associated with $w_c$ as a function of time ($T$) for different field strengths ($\mathcal{F}$) in the case of the O-band initial state.

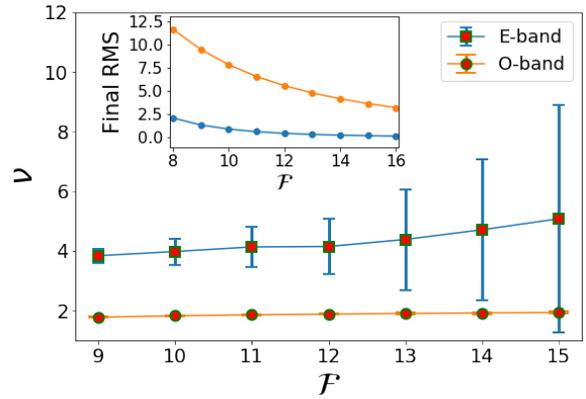

FIG. 5. Scaling exponents of the group velocity defined in Eq. (21) and calculated according to Eq. (22) plotted as function of field strength $\mathcal{F}$ for two types of initial states, which are approximately E- and O-band eigenstates. The error bars are the standard deviations calculated over time. The inset shows RMS at the final time of the simulation as a fucntion of the field strength for O- and E-band cases. The colors match the main figure.

similar results.

The exponent $\nu$ from Eq. (15) is calculated via

$$Rv_c \propto \frac{1}{\mathcal{F}^\nu} \Rightarrow \nu = \frac{\partial \log(Rv_c)}{\partial \log(\mathcal{F})} \quad (21)$$

that we discretize as:

$$\nu(T, \mathcal{F}) = \frac{\log[Rv_c(T, \mathcal{F} + 1)] - \log[Rv_c(T, \mathcal{F} - 1)]}{\log[\mathcal{F} + 1] - \log[\mathcal{F} - 1]}. \quad (22)$$

Figure 5 shows the scaling exponent $\nu$ (averaged over time $T$) as a function of field strength $\mathcal{F}$. Its values match the predicted values [Eq. (17)] 2 and 4 for the O- and E-band initial states, respectively, up to numerical errors. The error bars in the figure are the standard deviations calculated over time. The larger error bars of the exponent for the E-band case originate from the very small values of RMS—of the order of lattice spacing—as shown in the inset of Fig. 5.

## VII. DISCUSSION AND CONCLUSION

We have shown that Hubbard-like contact interaction induces particle transport in Wannier–Stark flatbands on a square lattice subject to a DC field. For two particles, we observe the motion of a bounded pair along the direction perpendicular to the field, while the motion along the field direction as well as relative motion is suppressed. This partial delocalization is possible since the lattice dimension is greater than one, in contrast to the case of many-body Wannier–Stark localization in one dimension [33–35]. Our results provide a way of controlling the direction of the particle transport by tuning the field direction. The scaling of the group velocity with the field strength as $v_g \sim (t/\mathcal{F})^\nu$ with $\nu \geq |x| + |y|$ was obtained for specific field directions, while the generic scaling remains a conjecture. Verifying and proving this conjecture for all field directions $(x, y)$, interactions $U$, and field strengths $t/\mathcal{F}$ is an interesting open question. Our results were derived for distinguishable particles, but they are applicable for spinless bosons as well. Our setup can be realized experimentally with state-of-art technologies like superconducting quantum processors [36] or cold atoms in tilted optical lattices [37] where the gravitational field can act as a DC field.

Another open question is to understand the effect of disorder or a magnetic field on such bounded motion of the particles and whether either can enhance the group velocity. Adding more particles, considering a finite density of particles, or adding spin degrees of freedom are also expected to yield interesting phenomena. Our results should be valid for any two-dimensional Bravais lattice of connectivity four (number of nearest neighbor hoppings). An immediate generalization is to extend these results to other Bravais lattices with different connectivities. In particular, in three dimensions we expect to observe a two-dimensional motion. Another interesting direction is to search for cases of emerging topological properties due to interaction that are otherwise absent [38], as was studied in the Refs. [29, 30].

## ACKNOWLEDGMENTS

This work was supported by the Institute for Basic Science, Project Code (Project No. IBS-R024-D1).

## Appendix A: Deriving the Wannier eigenbasis from the Bloch eigenbasis

The Bloch momentum eigenstates of a single particle were calculated in Ref. [21] for four-nearest neighbor hop-

ping in a square lattice, which read

$$|\psi(a, \kappa)\rangle = \sum_{\mu, \sigma \in \mathbb{Z}} \sum_{z, \eta \in \mathbb{Z}} J_\mu \left(-\frac{2t}{x\mathcal{F}}\right) J_\sigma \left(-\frac{2t}{y\mathcal{F}}\right) \times$$

$$\delta_{\mu x + \sigma y + z, a} e^{ik(\mu\tau_1 - \sigma\tau_2) + ik\eta} |z, w\rangle, \tag{A1}$$

where $a$ is the band index. The coordinate perpendicular to the field is $w = z(\tau_2 y - \tau_1 x) + \eta(x^2 + y^2)$. It is the sum of a (coordinate along the field) $z$-dependent part and an independent part. $\kappa$ is the single particle momentum conjugate to the independent perpendicular coordinate $\eta \in \mathbb{Z}$. The Wannier function is an eigenstate for the single particle case, and we construct a complete orthogonal set of spatially localized eigenstates by applying the inverse Fourier transform:

$$|\Phi(a, l)\rangle = \frac{1}{2\pi} \int_0^{2\pi} e^{-i\kappa l} |\psi(a, \kappa)\rangle \, d\kappa$$

$$= \sum_{n,m} J_{n-n_0} \left(\frac{2t}{x\mathcal{F}}\right) J_{m-m_0} \left(\frac{2t}{y\mathcal{F}}\right) |n, m\rangle, \tag{A2}$$

where $(n, m)$ are the lattice sites of a square lattice along conventional Cartesian axes, and the average location of the eigenstate $|\Phi(a, l)\rangle$ is controlled by two integers,

$$n_0 = a\tau_2 + ly, m_0 = a\tau_1 - lx. \tag{A3}$$

The Wannier eigenstates are parameterized by two indices: the band index $a = n_0 x + m_0 y$, and $l \in \mathbb{Z}$ which is the value of $\eta$ at coordinate $(n_0, m_0)$. In other words $(a, l)$ determines our coordinates for F-states/eigenstates. The $l$ takes countably infinite values and hence a eigenenergy $E = a\mathcal{F}$ is macroscopically degenerate.

## Appendix B: Geometric analysis for a different field direction

In Figs. 6 and 7 we depict the square lattice formed by F-sites for the DC field direction $(x, y) = (2, -3)$. Wwe denote single particle flatbands by straight lines over all F-sites that have the same band index $a$ or $z$ coordinate value for various $l$. In this representation, if an F-site has energy $a$, then from Eq. (4) we see that the nearest F-sites along the (horizontal) $n$-axis has energy $a \pm x$ and along the (vertical) $m$-axis has energy $a \pm y$. The number of single particle flatbands existing between two nearest F-sites is $|x| - 1$ along the $n$-axis, while this number is $|y| - 1$ along the $m$-axis.

### 1. $U = 0$ E-bands

Let us consider the two-particle Fock state located at F-site $(n_1, m_1, n_2, m_2)$. One possible E-band case is that both particles are located at the same F-sites $(n_1, m_1, n_2, m_2) = (n + \beta_e y, m - \beta_e x, n + \beta_e y, m - \beta_e x)$, which are parameterized by an integer $\beta_e$ for fixed $(n, m)$.

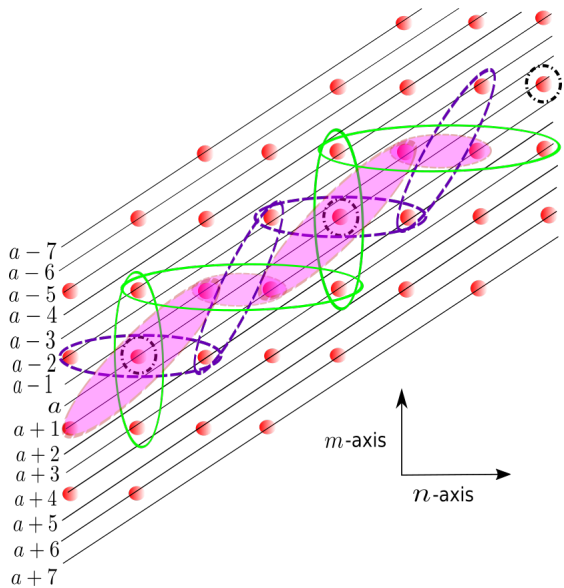

FIG. 6. Schematic representation of the eigenstate pairs for E-bands with eigenenergies $2a\mathcal{F}$ on the square lattice formed by F-sites for the DC field direction $(x, y) = (2, -3)$. Each of the lattice sites (red spheres) is a single particle F-state $|\Phi(a, l)\rangle$. $a \in \mathbb{Z}$ simultaneously denotes the single-particle band index and the $z$ coordinate. The constant $a$ lines are drawn over F-states $|\Phi(a, l)\rangle$ for different $l$, which characterize the average values of the perpendicular coordinate $\eta$ for fixed $a$.

This corresponds to the total eigenenergy $E^{(0)} = 2a\mathcal{F}$ (as $n_1x + m_1y + n_2x + m_2y = 2a$). Such states are shown in Fig. 6 by the F-sites at constant band $a$ surrounded by dash-dotted black circles. Two particles located at F-sites connected by reflection symmetry $(n + \zeta_e, m + \zeta'_e, n - \zeta_e, m - \zeta'_e)$, where $\zeta_e, \zeta'_e \in \mathbb{Z}$ but $\zeta_e$ and $\zeta'_e$ are not 0 simultaneously, correspond to the case where one particle is in the single-particle band $a + x\zeta_e + y\zeta'_e$ and the other particle is in the single particle band $a - x\zeta_e - y\zeta'_e$. Hence they have the eigenenergy $E^{(0)} = 2a\mathcal{F}$. In Fig. 6, we show three examples of such pairs: $(\zeta_e = 1, \zeta'_e = 0)$ by violet dotted ellipses connecting bands $a \pm 2$; $(\zeta_e = 0, \zeta'_e = 1)$ by green solid ellipses connecting bands $a \pm 3$; and $(\zeta_e = 1, \zeta'_e = 1)$ by magenta shaded ellipses connecting bands $a \pm 1$. Each of these pairs form a quasi one dimensional chain along the perpendicular direction of the field.

### 2. $U = 0$ O-bands

For the O-bands, two particles cannot occupy the same F-site. If $y$ is odd, two particles can form a pair if they are located at the nearest vertically oriented F-sites, and their eigenenergy is odd. If $x$ is odd, two particles can form such a pair if they are located at the nearest horizontally oriented F-sites. In Fig. 7, we show some examples of pair formation for odd eigen-ergy $E^{(0)} = (2a - 1)\mathcal{F}$. Pairs formed by particles located at bands $a + 1$ and $a - 2$ are shown within the orange ellipses by setting the location of the particles as $(n_1, m_1, n_2, m_2) = (n, m, n + \zeta_o, m + \zeta'_o)$. The location of such two-particle pairs nearest to $(n, m, n + \zeta_o, m + \zeta'_o)$ is at $(n, m, n + \zeta_o \pm y, m + \zeta'_o \mp x)$. Some other types of connected pairs are shown within cyan-shaded and blue-dashed ellipses. Similar to the E-band cases, they also form different types of quasi one dimensional chains along the perpendicular direction of the field.

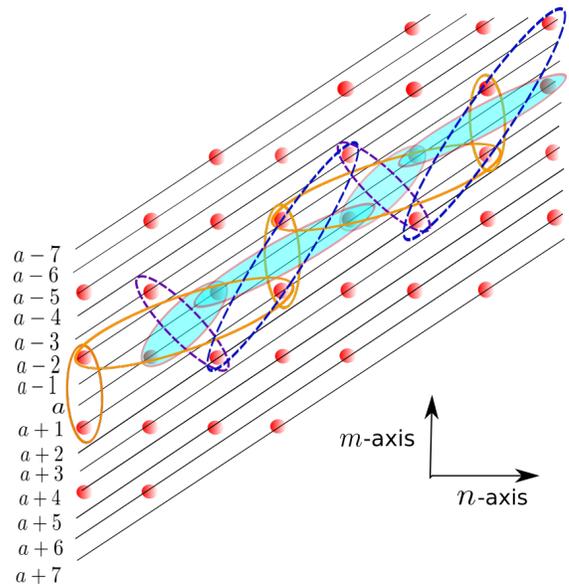

FIG. 7. Schematic representation of the eigenstate pairs for O-bands with eigenenergies $(2a - 1)\mathcal{F}$ on the square lattice of F-sites for the field direction $(x, y) = (2, -3)$. The setup is similar to Fig. 6.

For both types of bands, the pairs that lead to two-particle transport will contribute by at least the order $(t/\mathcal{F})^5$. This appears as the leading order contribution in the off-diagonal terms in matrix $\hat{V}^{(1)}$ [Eq. (14)].

### Appendix C: Relation between the RMS of c.m. coordinate $w_c$ and new coordinate $l_c$

The two-particle coordinates along the perpendicular direction to the DC field read

$$
w_c = \frac{w_1 + w_2}{2} = \frac{z_c(\tau_2 y - \tau_1 x) + (x^2 + y^2)(\eta_1 + \eta_2)}{2},
$$
$$
w_r = \frac{w_1 - w_2}{2} = \frac{z_r(\tau_2 y - \tau_1 x) + (x^2 + y^2)(\eta_1 - \eta_2)}{2}.
$$
(C1)

In our numerical analysis,

$$
\eta_1 + \eta_2 \equiv 2l_c + \xi, \eta_1 - \eta_2 \equiv \xi,
$$
$$
\tau_2 y - \tau_1 x = 1, (x^2 + y^2) = 2.
$$
(C2)

Therefore,

$$w_c = \frac{z_c}{2} + 2l_c + \frac{w_r}{2} - \frac{z_r}{2}. \tag{C3}$$

The averages of $w_c$ and $w_c^2$ are

$$\langle w_c \rangle = \frac{1}{2}\langle z_c + w_r - z_r \rangle + 2\langle l_c \rangle,$$
$$\langle w_c^2 \rangle = \frac{1}{4}\langle (z_c + w_r - z_r)^2 \rangle + 4\langle l_c^2 \rangle + 2\langle l_c(z_c + w_r - z_r) \rangle. \tag{C4}$$

Thus, the second moment of $w_c$ reads as follows:

$$\langle w_c^2 \rangle - \langle w_c \rangle^2 = \frac{1}{4}\left[ \langle (z_c + w_r - z_r)^2 \rangle - \langle z_c + w_r - z_r \rangle^2 \right]$$
$$+ 2\left[ \langle l_c(z_c + w_r - z_r) \rangle - \langle l_c \rangle\langle z_c + w_r - z_r \rangle \right]$$
$$+ 4\left[ \langle l_c^2 \rangle - \langle l_c \rangle^2 \right]. \tag{C5}$$

If $z_c$, $w_r$, $z_r$ are time independent, then we can write

$$\langle w_c \rangle = 2\langle l_c \rangle, \quad \langle w_c^2 \rangle - \langle w_c \rangle^2 = 4\left[ \langle l_c^2 \rangle - \langle l_c \rangle^2 \right]. \tag{C6}$$

and the RMS defined according to the physical distance in lattice spacing units is given by

$$Rw_c = \frac{\sqrt{\langle w_c^2 \rangle - \langle w_c \rangle^2}}{\sqrt{x^2 + y^2}} = \sqrt{2}\sqrt{\langle l_c^2 \rangle - \langle l_c \rangle^2}. \tag{C7}$$